# Chiral Magnonic Resonators:
# Rediscovering the Basic Magnetic Chirality in Magnonics

V. V. Kruglyak[a]

*University of Exeter, Stocker road, Exeter, EX4 4QL, United Kingdom*

## Abstract

The outlook for producing useful practical devices within the paradigm of magnonics rests on our ability to emit, control and detect coherent exchange spin waves on the nanoscale. Here, we argue that all these key functionalities can be delivered by chiral magnonic resonators – soft magnetic elements chirally coupled, via magneto-dipole interaction, to magnonic media nearby. Starting from the basic principles of chiral coupling, we outline how they could be used to construct devices and explore underpinning physics, ranging from basic logic gates to field programmable gate arrays, in-memory computing, and artificial neural networks, and extending from one- to two- and three-dimensional architectures.

[a] Corresponding author: V.V.Kruglyak@exeter.ac.uk

# I. Introduction

Spin waves are collective precessional modes of magnetically ordered materials. Due to the chirality of the magnetisation precession (Fig. 1 (a)), chiral and non-reciprocal phenomena should be abundant in magnonics, the study of spin waves.[1-3] Yet, in practice, they are scarcer than expected. For instance, let us take the magnonic dispersion relation, $f(\mathbf{k})$, where $f$ and $\mathbf{k}$ are the spin wave frequency and wave vector, respectively. Its non-reciprocity, $f(\mathbf{k}) \neq f(-\mathbf{k})$, is mainly associated with the Dzyaloshinskii–Moriya interaction (DMI).[4] However, the effect is relatively weak, so that it is often either neglected or even unnoticed whatsoever. We leave DMI and associated chiral phenomena outside of this piece, just noting a few recent works on the topic.[5-8] Of our interest are opportunities for design of spin wave devices that arise from the use of the more basic, more general and, arguably, more robust kind of magnetic chirality – that of the magneto-dipole field associated with precessing magnetisation. This chirality is well known and routinely exploited for spin waves in the so-called Damon-Eshbach spin wave (DESW) geometry: the magnonic wave vector is orthogonal to the direction of the in-plane static magnetisation (Figs. 1 (b,c)).[9] In films, structures and more generally planar magnonic media, DESWs have surface character – their amplitude is distributed asymmetrically relative to the medium's horizontal symmetry plane. When either the direction of propagation or the magnetisation is reversed, this distribution is also inverted relative to this plane. By breaking the reflection symmetry of the medium,[10] the amplitude asymmetry can be converted into non-reciprocity of the DESW dispersion relation.[11-14] Yet, there are both practical and fundamental reasons to widen the search for chiral spin wave phenomena and devices beyond the DESW geometry. For instance, the backward volume spin wave (BVSW) geometry (the wave vector is parallel to the in-plane static magnetisation, Fig. 1 (d)) suits best bias free architectures containing continuous magnonic waveguides magnetised by their shape anisotropy.[15-17] The forward volume spin wave (FVSW) geometry (the magnetisation is orthogonal to the film plane, Fig. 1 (e)) is essential if isotropy of the dispersion relation is required.[18-20] In perceived two-dimensional (2D) and three-dimensional (3D) magnonic architectures,[21] sticking to a particular spin wave geometry might become impractical at all.

As far as spin wave devices are concerned,[22] it is straightforward to convert non-reciprocity of the dispersion relation into a non-reciprocal modulation of the spin wave phase $\varphi = kd$, since $k(f, \widehat{\mathbf{k}}) \neq k(f, -\widehat{\mathbf{k}})$, where $\widehat{\mathbf{k}}$ and $k$ are the unit vector along and absolute value of the wave vector, respectively. However, the distance travelled, $d$, is likely to be large. Furthermore, one might need to rely on magnetic damping and non-reciprocity of the group velocity, $\mathbf{v}_\mathrm{g}(f, \widehat{\mathbf{k}})$, (or perhaps, additional interference) to design e.g. a spin wave diode.[23] Hence, such devices would be difficult to downscale, e.g. to make smaller than a few spin wave wavelengths at the relevant frequency. A superior means of control could be through the spin wave amplitude and phase variation induced by scattering. However, a non-reciprocal dispersion relation does not readily translate into non-reciprocal scattering parameters (i.e. transmission and reflection coefficients) of spin waves.[24]

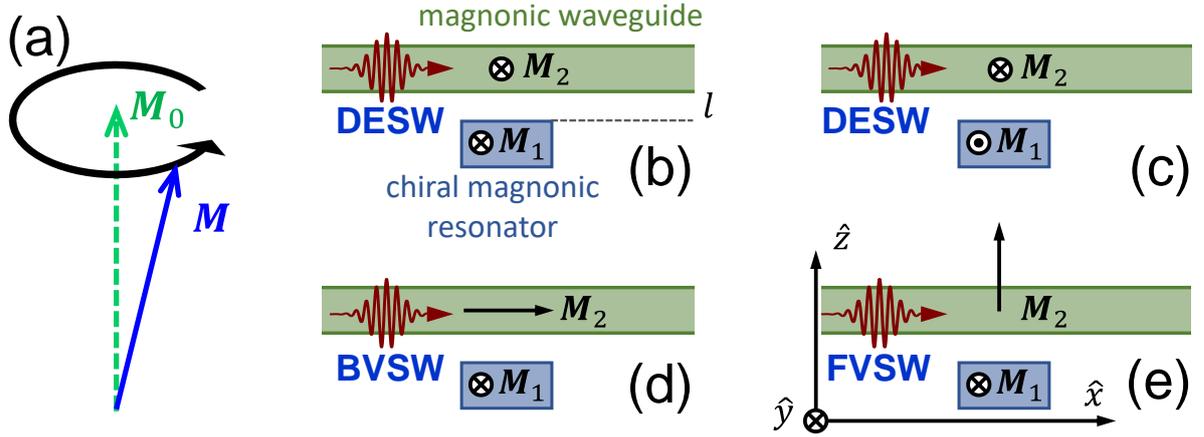

Fig. 1   Panel (a) shows Larmor precession of the magnetisation vector $\boldsymbol{M}$ around its equilibrium value $\boldsymbol{M}_0$. Panels (b) – (e) illustrate the basic chiral magnetic configurations of the magnonic waveguide – resonator system. In each case, we show the system's cross-section by the "sagittal plane" – the $(x, z)$ plane in our case. This plane is parallel to the spin wave's wave vector and normal to the substrate, which is assumed to be parallel to the $(x, y)$ plane. The resonator's magnetisation, $\boldsymbol{M}_1$, points in the $\hat{\boldsymbol{y}}$ direction in panels (b), (d) and (e), and in $-\hat{\boldsymbol{y}}$ direction in panel (c). The waveguide's magnetisation, $\boldsymbol{M}_2$, points in the $\hat{\boldsymbol{y}}$ direction in panels (b) and (c) (DESW geometry), in the $\hat{\boldsymbol{x}}$ direction in panel (d) (BVSW geometry), and in the $\hat{\boldsymbol{z}}$ direction in panel (e) (FVSW geometry).

The brief discussion above explains the significance of the demonstration of the basic chiral functionalities (devices) – unidirectional emission (source or input coupler), unidirectional phase inversion (phase-shifter), unidirectional transmission (diode / valve), and unidirectional absorption (output coupler / trap) – achieved in Refs. 15,16 for travelling spin waves resonantly coupled via their stray magnetic field to neighbouring subwavelength magnonic resonators (Fig. 1). The chirality of the devices is based on the chirality of the magnetic precession in the resonators, which makes them re-programmable by switching their magnetisation. With variations, the findings from Refs. 15,16 are gaining recognition as the state of the art and possibly the future of spin wave excitation, manipulation, and detection on the nanoscale. Hence, this Perspective focuses upon and elucidates recent developments based on the same physics, and then uses obtained insights to lay out a roadmap for further progress in the areas of magnonics and magnonic technology associated with the use of chiral magnonic resonators.

**II. Chiral Magneto-Dipole Coupling and Where to Find It**

Let us begin by making a general remark. The chirality of the magnetic precession plays a role only in its interaction with a chiral stimulus, e.g. a circularly polarised dynamic magnetic field. The interaction is strong when the chiralities (senses of rotation) match and is weak or even absent otherwise. The chirality of the magnetic precession is ubiquitous. The chirality of the dynamic magneto-dipole field ("stray field") can be traced back to the magnetic potential satisfying the Laplace equation[25] and so is ubiquitous too! With this in mind, the origin of the chiral coupling between a resonator's modes and propagating spin waves can be considered in two ways: (i) by considering the Zeeman energy of spin waves' dynamic magnetization in the resonator's dynamic stray field, and (ii) by considering the Zeeman energy of the resonator in the spin waves' stray field. Due to the magnetostatic reciprocity theorem [26,27], the two energies are equal, and the two approaches are therefore equivalent. Here, we focus on approach (i), which was implemented numerically in Refs. 15,25 (Fig. 2).

Figure 2 illustrates chirality of the dynamic stray field of a resonator magnetised in the positive $y$ direction, i.e. orthogonal to the sagittal plane.[15] In the real space, one notices a "pattern of the dipolar field that starts from the right and moves towards left from (a) to (d), echoing a wave travelling towards negative $x$ direction" [15] just below the resonator. Not so clear in Fig. 2, a similar pattern propagates in the opposite direction just above the resonator. The direction in which the pattern propagates above and below the resonator is determined by the sense of rotation of the dynamic magnetisation in the resonator, i.e. by the chirality of its Larmor precession (Fig. 1 (a)). If we switch the direction of the static magnetisation from the positive to negative $y$ direction, the sense of rotation will change from clockwise to anticlockwise, i.e. its polarisation will switch from right- to left-circular. As a result, the travelling patterns will reverse their propagation directions.

In the reciprocal space, these travelling patterns lead to asymmetry of the Fourier spectra relative to line $k_x = 0$: the Fourier amplitude distributions are shifted to the left half ($k_x < 0$) of the reciprocal space below resonator, as shown in Fig. 2 (e), and to the right half ($k_x > 0$) above the resonator (not shown). The Zeeman coupling energy of the resonator's field to a propagating spin wave mode is proportional to the field's Fourier amplitude for a wave number coinciding, in both magnitude and sign, with that of the propagating mode. Hence, the asymmetry of the Fourier spectra shown in Fig. 2 (e) leads to a difference in the coupling energy of the resonator's field to waves with opposite signs of $k_x$. This asymmetry is reversed relative to the $k_x = 0$ axis when the magnetization of the resonator and therefore the chirality of its precession are reversed.[15]

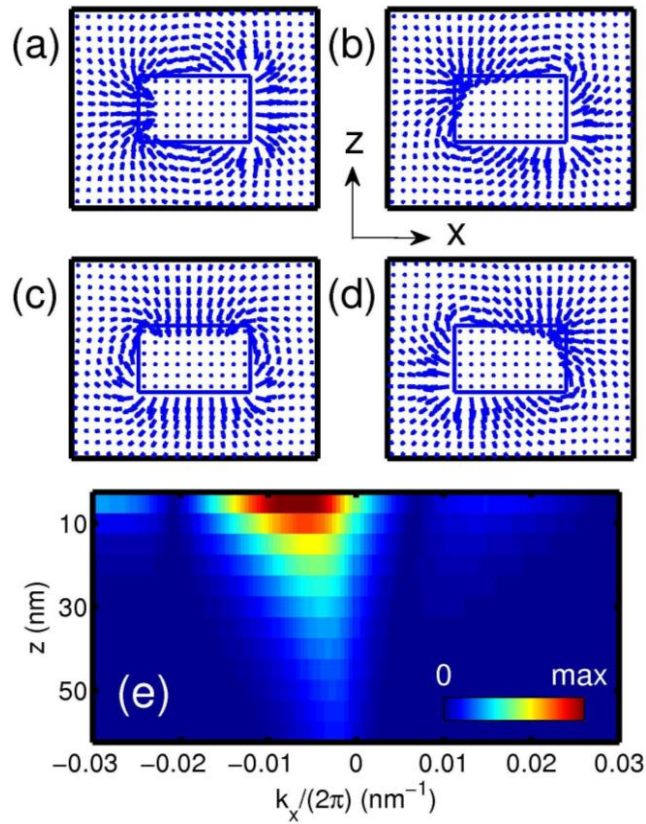

Fig. 2    Panels (a) to (d) show snapshots of the dynamic stray magnetic field of the resonator in the sagittal plane at relative simulation times of 0, 1/8, 1/4, and 3/8 of the precession period, respectively. (e) The spatial Fourier transform along the $x$ direction of the field's $z$ component is shown for different vertical distances (labelled as '$z$') from the lower surface of the resonator. Reproduced from Ref. 15, with permission of AIP Publishing.

The discussion above both demonstrates and explains fully the chirality of the coupling of the resonator's field to spin waves propagating in the magnonic waveguide without any reference to the direction of its static magnetisation relative to the spin waves' wave vector. In the BVSW geometry (Fig. 1 (d)), considered Refs. 15,16,25, only the $z$ component of the resonator's dynamic field contributes to the Zeeman coupling energy and is used in the analysis in Fig. 2 (e). In contrast, the $x$ component of the resonator's field is parallel to and therefore does not exert any torque upon the waveguide's static magnetisation. From the energy viewpoint, the field's $x$ component is orthogonal to the spin waves' magnetisation and therefore does not contribute to the Zeeman coupling energy.

In a FVSW geometry (Fig. 1 (e)), the $x$ and $z$ components of the resonator's field would switch their roles. In the DESW geometry (Fig. 1 (b-c)), both the $x$ and $z$ components matter. Moreover, it becomes essential that they vary with a 90° phase shift relative to each other, i.e. that the resonator's dynamic stray field is actually circularly polarized in the sagittal plane. The helicity of the circular polarisation is opposite to that of the resonator's dynamic magnetisation and is therefore determined by the direction of the static magnetization. The dynamic magnetisation of spin waves in the DESW geometry is elliptically polarised in the sagittal plane. When the static magnetisations in the resonator and waveguide are antiparallel ("antiparallel DESW geometry", Fig. 1 (c)), the sense of rotation of the resonator's stray field and the waveguide's dynamic magnetisation coincide, and the energy of their interaction is maximised, i.e. the coupling is strong. In contrast, the coupling is minimised when the resonator and waveguide are magnetised parallel to each other ("parallel DESW geometry", Fig. 1 (b)). Hence, the chirality is enhanced (suppressed) in the antiparallel (parallel) DESW geometry relative to that in the BVSW and FVSW geometries.

Expectedly, approach (ii) confirms that, fundamentally, the DESW, BVSW and FVSW geometries may all host chiral magnonic phenomena and underpin practical chiral spin wave devices, provided one knows where to look. So, we just outline this approach, a systematic theoretical discussion of which can be found e.g. in Refs. 28-30. Whether you consider DESWs, BVSWs or FVSWs, their stray fields are circularly polarised in the 'sagittal plane'. The senses of the field's rotation are opposite on opposite sides of the film, and the sides switch when the propagation direction is reversed. So, the coupling energy of the field and the resonator's magnetisation is enhanced on the side where the senses of the field's and magnetisation's rotations match and is suppressed otherwise. The DESW geometry stands out only in that the stray field on the opposite sides has different magnitudes, in addition to the opposite helicities.

Using either approach, and assuming that the dynamic magnetisations are uniform both in the resonator and in the cross-section of the magnonic waveguide, one can write for the Zeeman energy of their interaction

$$E_Z \propto e_w e_r = m_{w,z} m_{r,z}(1 - \varepsilon_w \varepsilon_r + \sigma(\varepsilon_r - \varepsilon_w)\,\text{sign}(k_x)), \qquad (1)$$

where $m_{w(r),x}$ and $m_{w(r),z}$ are the projections of the dynamic magnetisation in the waveguide (resonator) onto the sagittal plane, $\varepsilon_{w(r)}$ are the corresponding ellipticities defined by $m_{w(r),x} = i\varepsilon_{w(r)} m_{w(r),z}$, and $\sigma = +1$ when the resonator is below the waveguide (as in Fig. 1) and $\sigma = -1$ when it is above the waveguide (as in Ref. 15). The factor $e_w = m_{w,z}(1 - \sigma \varepsilon_w \,\text{sign}(k_x))$ is specific to the magnonic waveguide, describing chirality of its interaction with evanescent field of any origin, e.g. as in the case of chiral excitation of spin waves in the DESW geometry by a dynamic magnetic field from a nonmagnetic microwave stripline.[30-32] This factor is chiral unless either $m_{w,x} \equiv 0$ (BVSW geometry) or $m_{w,z} \equiv 0$ (FVSW geometry). The factor $e_r = m_{r,z}(1 + \sigma \varepsilon_r \text{sign}(k_x))$ describes chirality associated with the dipolar nature of the source of the magnetic field,[25,30] i.e. with the fact that the field originates from the dipole moment of the resonator's dynamic magnetisation. This factor is chiral unless either $m_{r,x} \equiv 0$ or $m_{r,z} \equiv 0$. For the DESW geometry, the chirality in Eq. (1) vanishes when the dynamic magnetisations of the waveguide and resonator rotate in the same sense and with equal ellipticities, i.e. when $\varepsilon_r = \varepsilon_w$. The chirality is perfect (i.e. the coupling vanishes completely for one sign of $k_x$) when the precession is circularly polarised either in the resonator or in the waveguide (i.e. when either $|\varepsilon_w| = 1$, or $|\varepsilon_r| = 1$). The chirality is strongest (i.e. $|E_Z(k_x) - E_Z(-k_x)|$ is maximised) when $\varepsilon_r = -\varepsilon_w = \pm 1$, which corresponds to the DESW geometry with antiparallel magnetisations of the resonator and waveguide (Fig. 1 (c)).

The main missing ingredient of Eq. (1) is the 'form factor'. The latter describes how the finite dimensions of the resonator (which may be driven by external dynamic stimuli) affect coupling of its modes (and by extension, of the external stimuli) to propagating spin waves. This form factor is not chiral. Examples of its calculation can be found e.g. in Ref. 16 and 28 for uniform modes of resonators with cylindrical and rectangular cross-sections, respectively. The resonator's dimensions, via the form factor, also define the dependence of the coupling strength on its distance from the waveguide.

**III. Why Chiral Resonant Magnonic Devices?**

To illustrate practical benefits of chirality, let us consider a resonant magnonic diode (also known as 'isolator' or 'valve') – a device that transmits spin waves incident from one direction and blocking counter-propagating waves, using the simple phenomenological model from Ref. 25 (Fig. 3 (a)). The model postulates existence of chiral coupling between a mode, $\varphi(t)$, of a point-like magnonic resonator and spin waves propagating to the right, $\psi_R(t, x)$ (left, $\psi_L(t, x)$), with complex coupling strengths, $\Delta_{R(L)}$. Then, the resonator's mode as well as the amplitude transmission, $\tau_{R(L)}$, and reflection, $r_{R(L)}$, coefficients can be calculated for the right (left) propagating spin waves as

$$\varphi_{R(L)} = \frac{\Delta^*_{R(L)}}{\omega - \Omega_0 + i\Gamma_{tot}}, \quad \tau_{R(L)} = \frac{\omega - \Omega_0 + i(\Gamma_0 - \Gamma_{R(L)} + \Gamma_{L(R)})}{\omega - \Omega_0 + i\Gamma_{tot}}, \quad r_{R(L)} = -\frac{i}{v}\frac{\Delta^*_{R(L)}\Delta_{L(R)}}{\omega - \Omega_0 + i\Gamma_{tot}}, \qquad (2)$$

where $\Gamma_{R(L)} \equiv |\Delta_{R(L)}|^2/(2v)$ are the radiative linewidths into the right (left) propagating plane waves, $\Gamma_{tot} = \Gamma_0 + \Gamma_R + \Gamma_L$ is the total linewidth of the resonator, $\Gamma_0$ and $\Omega_0$ are the resonator's dissipative linewidth and angular frequency, $\omega$ and $v$ are the angular frequency and group velocity of the propagating modes, and the subscript in $\varphi_{R(L)}$ corresponds to that of the incident propagating wave.

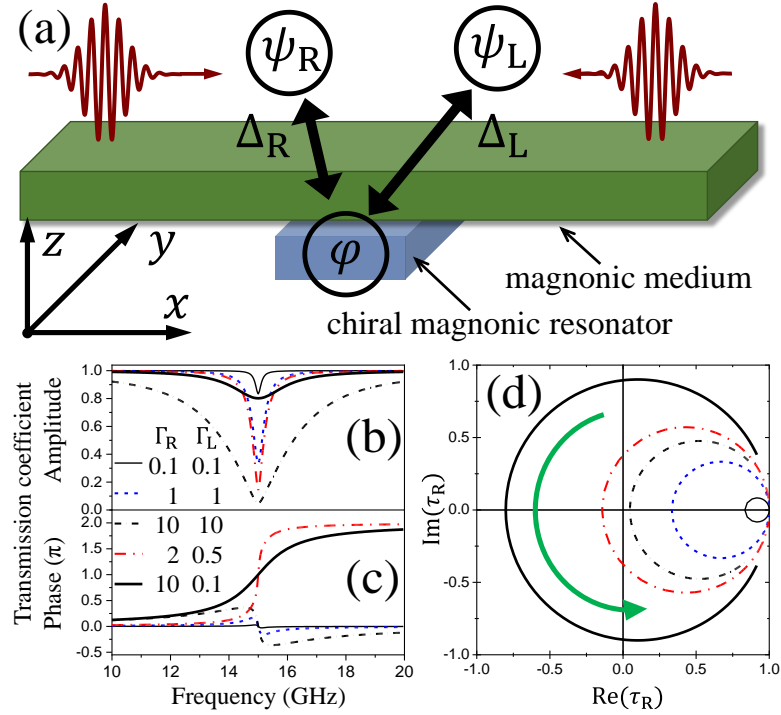

Fig. 3  Panel (a) illustrates the phenomenological model: the magnonic resonator hosts mode $\varphi$ that couples chirally to spin waves propagating in the magnonic medium to the left (mode $\psi_L$) and to the right (mode $\psi_R$), with coupling strengths of $\Delta_L$ and $\Delta_R$, respectively. Panels (b) and (c) show the magnitude and phase, respectively, of the transmission coefficient of the right going waves given by Eq. (2) as a function of their frequency, $f = \omega/2\pi$. The values of the radiative linewidths $\Gamma_R$ and $\Gamma_L$ are given as fractions of the resonator's dissipative linewidth $\Gamma_0 = 2\pi \times 0.5$ GHz, while its resonance angular frequency is $\Omega_0 = 2\pi \times 15$ GHz. (d) The dependences from (b) and (c) are plotted in the complex plane, using the frequency as a parameter with the arrow indicating the direction of its increase. Reproduced with permission from Ref. 25. Copyright 2021 American Physical Society Publishing.

First of all, we note that chirality, $\Gamma_R \neq \Gamma_L$, is a prerequisite for the diode function. Less trivial is the chirality-induced enhancement of the amplitude modulation as a function of the frequency. Figure 3 presents the frequency dependence of the complex transmission coefficient calculated using Eq. (2). The minimum transmission (i.e. maximum modulation) is achieved at resonance, $\omega = \Omega_0$. For a reciprocal (non-chiral, $\Gamma_R = \Gamma_L$) resonator, the minimum is $\Gamma_0/\Gamma_{tot}$, while for a chiral resonator, the transmitted wave can vanish altogether if condition $\Gamma_R = \Gamma_0 + \Gamma_L$ is met. The latter condition may be fulfilled even for resonators with relatively high damping (e.g. as in metallic ferromagnets) and quite significant spacings ($l$ in Fig. 1 (b)) between the magnonic resonator and waveguide.[15,25]

The use of the system as a phase shifter benefits from its chirality even more strongly (Fig. 3 (c,d)). The phase inversion, i.e. a phase shift of $\pi$, at resonance and arbitrary phase shifts, from 0 to $2\pi$, around the resonance are easily obtained if $\Gamma_R > \Gamma_0 + \Gamma_L$. However, the amplitude modulation is undesirable in this case. The parasitic reduction of the transmitted amplitude is $1 - |\tau_R(\omega = \Omega_0)| = 2/\left(1 + \frac{\Gamma_R}{\Gamma_0 + \Gamma_L}\right)$ can be minimised by simultaneously suppressing both the dissipative and backscattering radiative linewidths. For a non-chiral resonator ($\Gamma_R = \Gamma_L$), the achievable phase shifts always remain small, since the "loop" representing the transmission coefficient in the complex plane (Fig. 3 (d)) never contains the origin.

Equations (2) predict that the reflectance, $R_{R(L)} = |r_{R(L)}|^2$, from a strongly chiral resonator is both weak and reciprocal, which is due to $r_R \propto \Delta_R^* \Delta_L$, where $|\Delta_R^*| \gg |\Delta_L|$. Hence, the chirality promotes the absorption of the incident spin wave power compared to its reflection. Indeed, defining the absorbance, $A_R$, as

$$A_R = 1 - |\tau_{R(L)}|^2 - |r_{R(L)}|^2 = \frac{4\Gamma_R \Gamma_0}{(\omega - \Omega_0)^2 + \Gamma_{tot}^2}, \quad (3)$$

we find that $A_R/R_R = \Gamma_0/\Gamma_L$, i.e. the absorbance to reflectance ratio increases as the resonator's coupling to the counter-propagating wave decreases. The maximum of the resonant absorption corresponds to the same condition $\Gamma_R = \Gamma_0 + \Gamma_L$ as that for zero resonant transmission, and it can reach unity if $\Gamma_L = 0$. In this latter case, the incident wave is fully "trapped"[15,33] – another effect possible only when the resonator is chiral. This wave trapping is obviously non-reciprocal, which is essential for reading out the pseudospin within the paradigm of "magnon valleytronics".[34]

Assuming that the resonator's modes are driven by another stimulus, e.g. by an external microwave magnetic field as in Ref. 16, it will emit spin waves into the waveguide, with their amplitudes given by

$$c_{R(L)} = -\frac{i\Delta_{R(L)}}{v}\varphi. \quad (4)$$

The amplitudes of the spin waves emitted to the right, $c_R$, and to the left, $c_L$, differ by a factor equal to the ratio of the coupling coefficients, $\Delta_R/\Delta_L$, another property of possible practical use. So, the resonator acts as a chiral microwave-to-spin-wave transducer,[16] i.e. a device concentrating the power of incident microwaves[35] and converting it into unidirectional propagating spin waves.[36]

The emission of spin waves leads to broadening of the mode's resonance. So, assuming that it is driven by an incident microwave magnetic field that couples to the mode with coefficient $\Delta$, the mode is given by

$$\varphi = \frac{\Delta h}{\omega - \Omega_0 + i\Gamma_{\text{tot}}}. \tag{5}$$

The resonant enhancement by a factor of $(\omega - \Omega_0 + i\Gamma_{\text{tot}})^{-1}$ is a characteristic feature of the physics above. Together with the chiral resonance in the reciprocal space described in the previous section, this resonance justifies the associated effects and devices to be called "resonant".

## IV. Chiral Magnonic Resonators as Building Blocks of the Generic Magnonic Device

We now discuss how chiral magnonic resonators can be used to build the spin wave input coupler (source), output coupler (detector), and control element of the generic magnonic device from Ref. 1 (Fig. 4). The use of the resonators as sources and detectors is based on their ability to mediate coupling between the external electrical circuitry and spin waves. Indeed, the quasi-uniform precessional modes in the resonators can couple both to propagating short-wavelength spin waves and to nearly uniform microwave magnetic fields, as shown in Figs. 5 and 6.[15,16] This transduction functionality is not limited to inductive coupling and will also be useful when coupling magnonic devices e.g. to electric fields[37] and currents.[38] Back to the inductive coupling, the micrometre-scale experimental demonstration of a spin wave source in Ref. 16 was followed by Ref. 39, where not only unidirectional emission but also detection of spin waves with wavelengths down to 48 nm and frequencies in excess of 20 GHz was achieved using individual nano-sized chiral magnonic resonators. This both proves the viability of the advocated here approach to construction of spin wave devices and sets the size benchmark for magnonic devices in general.

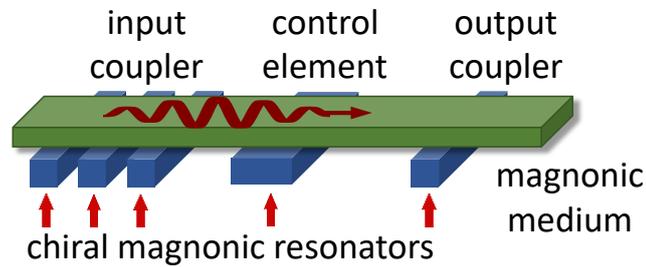

Fig. 4  The generic magnonic devices from Ref. 1 implemented using chiral magnonic resonators. The uniform modes in the input and output resonators are inductively coupled to the external microwave circuitry (not shown). The dark mode in the wider control resonator does not couple to the external microwave fields but can affect the phase and amplitude of spin waves propagating in the magnonic medium.

One may choose to arrange chiral magnonic resonators into periodic arrays, proposed for spin wave emission already in Ref. 16 (see the inset in Fig. 5) and investigated as "grating couplers" e.g. in Refs. 29,40-42. However, benefits of using such arrays for spin wave emission and detection are neither obvious nor guaranteed. Combining Eqs. (4) and (5), one can find that the condition for strongest resonant emission from a single field-driven resonator coincides with that for zero resonant transmission (maximal resonant absorption) of spin waves propagating in the same direction, e.g. $\Gamma_R = \Gamma_0 + \Gamma_L$ for right going waves. This means, for example, that the spin wave emitted to the right by a finite array of resonators each satisfying this condition and driven at resonance, i.e. $\omega = \Omega_0$, is due to the rightmost resonator only. Spin waves emitted to the right by the other resonators are either absorbed within the array or back-reflected to the left, which is the direction of weaker emission from a stand-alone resonator (since $\Gamma_R > \Gamma_L$). Then, the spin wave emitted by the array to the left is enhanced as a result of multiple reflections and interference of waves emitted by all its resonators. Similar cumulative enhancement of emission in a particular direction occurs whenever $\tau_{R(L)} \neq 0$, i.e. the array in question behaves as a grating coupler. This includes off-resonance emission, i.e. $\omega \neq \Omega_0$, and arrays of non-chiral resonators, i.e. $\Gamma_R = \Gamma_L$.

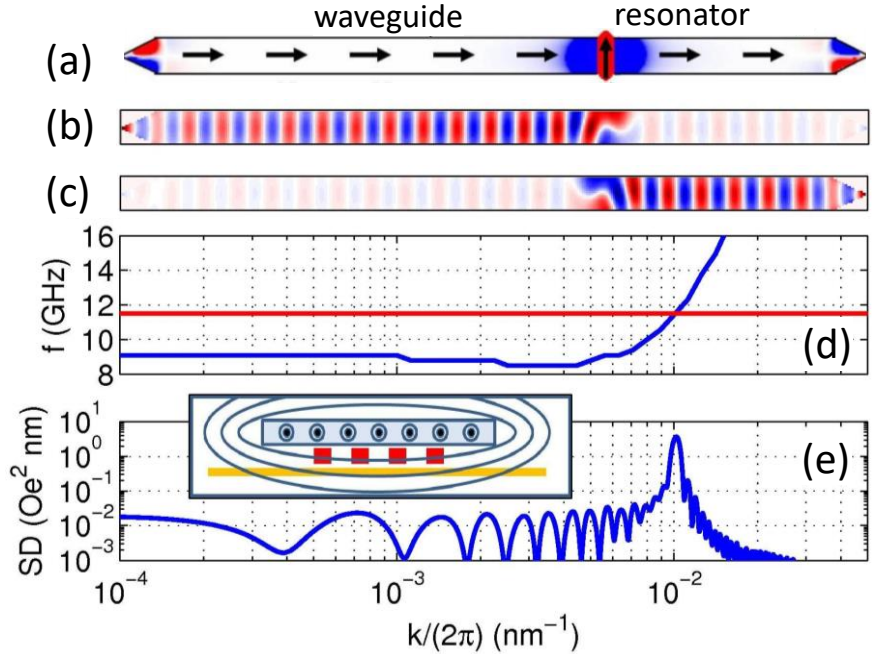

Fig. 5 Chiral magnonic resonator as a unidirectional spin wave source. Panel (a) shows the ground state of the system. The resonator is 50 nm wide, 150 nm long and 30 nm thick, and is separated by 10 nm spacing from the 10 nm thick waveguide. There is no bias field applied. The arrows represent the static orientation of the magnetisation. Panel (b) shows the spin wave emitted into the waveguide when the system from (a) is driven by a uniform microwave magnetic field at the quasi-uniform mode frequency of the resonator. Panel (c) shows the same as (b) for the case when the resonator is magnetised in the direction

opposite to that shown in (a); the direction of spin wave emission is switched as a result. (d) The dispersion relation of spin waves in the magnonic waveguide (blue) is shown together with the line (red) representing the quasi-uniform mode frequency of the resonator. (e) The spectral density of the emitted spin waves is shown. The inset shows the cross section of a microstrip carrying microwave current (running out of page) leading to the microwave magnetic field represented by elliptical lines. The microstrip is relatively wide, to reduce Ohmic losses. The microwave field excites an array of chiral magnonic resonators (red squares), which mediate the field's coupling to spin waves propagating in the magnonic waveguide (yellow slab). Reproduced (with modifications) from Ref. 16, with permission of AIP Publishing.

To quantify the emission enhancement achievable for arrays of chiral resonators, we consider the case of perfect chirality, when $\Gamma_L = 0$. Then, reflection from individual resonators vanishes altogether, and the amplitude of the spin wave emitted to the right by an array of $N$ resonators is given by

$$c_{R,N} = c_R(1 + \alpha\tau_R + (\alpha\tau_R)^2 + \cdots + (\alpha\tau_R)^N) = c_R \frac{1-(\alpha\tau_R)^N}{1-\alpha\tau_R}, \tag{6}$$

where the factor of $\alpha$ ($|\alpha| \leq 1$) accounts for the amplitude and phase changes acquired when the wave travels between neighbouring resonators. Eq. (6) shows that, for given $N$, the maximum enhancement is achieved when $\alpha\tau_R \cong |\alpha\tau_R|$, while little is gained when either $|\alpha|$ or $|\tau_R|$ is significantly smaller than unity. Recalling that the frustrated transmission is a necessary attribute of enhanced emission from individual chiral magnonic resonators, one can appreciate the problem with combining them into arrays: a stronger emission of spin waves comes at the expense of a greater proportion of power dissipated within the array. Hence, although the possibly stronger overall emission from grating couplers is of obvious benefit for research, the need for greater energy efficiency and smaller footprint favours individual resonators for use in practical devices.

For detection, stand-alone chiral magnonic resonators (see Fig. 6) designed to meet the condition of optimal resonant absorption (zero resonant transmission) should outperform arrays at resonance. Off resonance, when transmission is non-zero, additional resonators could help to harvest the incident power more fully, again at the expense of a larger footprint. Generally, the issue of bandwidth naturally arises for resonant devices. The discussion above suggests that the use of arrays could broaden the bandwidth but only somewhat: the reliance on the frequency resonance in individual resonators would be weakened but another restriction would emerge instead – that of the Bragg resonance.[29,40-42]

Before discussing the control resonator from Fig. 4, we note that it is actually optional. Indeed, a great deal of control can be achieved already using the input and output resonators, e.g.

by switching their magnetisations, as in the case of the magnonic NAND gates from Refs. 43,44. In other devices, such as the magnonic XNOR gate from Refs. 43,45, the control resonator is a necessary element. An experimental realisation of the spin wave amplitude control using sub-micrometre Fabry-Perot magnonic resonators was presented in Ref. 46. Their principle of operation is a somewhat different, being based on the nonreciprocity of the DESW geometry (Fig. 1 (b),(c)), while the spin waves in Ref. 46 had frequencies sizably smaller than those in Ref. 39. Yet, Ref. 46 demonstrates boldly the technological feasibility of the approach.

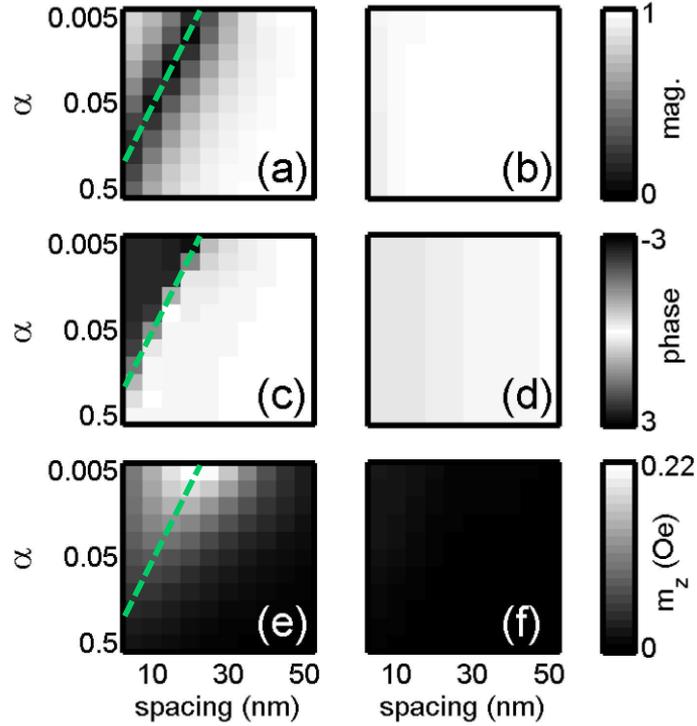

Fig. 6 Chiral magnonic resonator as a spin-wave control element and output coupler. The grey scale in panels (a) and (c) shows the relative magnitude and phase, respectively, of the spin wave transmitted from right to left underneath the resonator shown in Fig. 5 (a) as a function of the resonator-to-waveguide spacing and the resonator's Gilbert damping constant, $\alpha$. The spin wave is partly absorbed by the resonator, making it to precess. The volume-averaged magnitude of this precession is shown in (e). The green dashed lines are guides for the eye: they are the same in every panel and approximately correspond to the condition of minimum transmission (maximum absorption) discussed in the text. Panels (b), (d), and (f) show the same quantities as panels (a), (c), and (e), respectively, but for the opposite direction of the resonator's magnetisation. Reproduced (with modifications) from Ref. 15, with permission of AIP Publishing.

Taken together, measurements from Refs. 39,46 prove practical feasibility and scalability of the device architecture shown in Fig. 4. This is enforced by another aspect of the findings presented in Ref. 46 – the possibility of spin wave control using higher order modes in the resonator. As explained in Ref. 25, some of these are so called "dark modes" – modes that do not couple to uniform external stimuli, typically because they have antisymmetric profile. This minimises the control resonator's direct inductive coupling to the input and output fields. However, the dark modes are at least as good as the quasi-uniform modes for the chiral control of propagating spin waves.[25] Finally, we note that, in principle, one could consider replacing any of the resonators in Fig. 4 by alternative magnonic elements performing either the same or different function.[47-50] Such substitutions and their results are however beyond the scope of this discussion.

**V. Chiral Magnonic Resonators: A Brief Research Roadmap**

Knowing the physics presented in Refs. 15,16,25-30,33 and reassured by the impressive experimental demonstrations from Refs. 39,46, we can lay out the roadmap for further development of chiral magnonic resonators and devices, which could establish the associated sub-area of magnonics as a pervasive and versatile technology for data and signal processing. This brief roadmap complements to the most recent magnonics roadmap[2] and review,[3] which however do not cover the topic of chiral magnonic resonators.

**A. Basic studies and design of linear chiral magnonic resonators and devices in 1D**

The most immediate goals concern experimental realisation of devices and effects proposed theoretically but not yet (fully) achieved in practice. The resonators in the goals listed below should have dimensions of 100 nm (or smaller) and operate at frequencies of 10 GHz (or higher): after all, we have seen that the chirality is more pronounced and the devices operate better on the nanoscale. As in the existing studies, yttrium iron garnet (YIG)[51] is the material of choice (being de facto "silicon of magnonics") for thin film media[52,53] and conduits.[54-56] Co, Ni, Fe, and their alloys[39,46] offer a good range of saturation magnetisation and Gilbert damping values, required to tune the radiative and dissipative linewidths of the resonators. Hence, the task is to use the available materials to implement the earlier described design principles in practical devices.

A1. *Optimisation of spin wave emission and absorption (detection) through geometry tuning.* This geometrical fine-tuning (e.g. by changing the resonator's shape and dimensions and the resonator-to-medium spacing[15,28]) of the chiral coupling strength and therefore of the spin wave emission and absorption (detection), as discussed in Section III, remains an untapped resource.

A2. *Chiral phase shift and inversion.* This crucial functionality has not been addressed in Refs. 39,46 or elsewhere. The key is to ensure that $\Gamma_\text{R} \gg \Gamma_0 + \Gamma_\text{L}$. Yet, it seems the authors of Ref. 46 might just need to broaden the bandwidth of their measurements in the antiparallel DESW geometry, to achieve this. This is because the resonator's mode frequency is also affected by the coupling: not mentioned in Section III and being a relatively weak effect in the BVSW geometry,[25] this frequency shift turns out to be very large in the DESW geometry, which was used in Ref. 46.

A3. *Chiral magnonic resonators beyond the DESW geometry.* Micromagnetic simulations from Refs. 15,16,25 and experiments from Ref. 29 demonstrate feasibility of this task in the BVSW geometry, and the effects could then be optimised further. As discussed earlier, the FVSW geometry should exhibit equally strong effects.

A4. *Dark-mode resonators.* The observation of the Fabry-Perot resonance in Ref. 46 and micromagnetic simulations from Refs. 25,57 imply feasibility of using magnonic dark modes in resonators to construct devices. Now, one needs to downscale, experimentally, the resonator width, to demonstrate the chiral coupling of its lowest lying antisymmetric mode to propagating spin waves and absence of its coupling to uniform stimuli, and then to implement and optimise the resonators as e.g. magnonic diodes and phase shifters within the generic device shown in Fig. 4.

A5. *Complete chiral magnonic gate.* The aim is to demonstrate practically the complete device shown in Fig. 4 with all its constituent resonators individually optimised for their functions.

A6. *Resonators on laterally confined magnonic waveguides.* A common feature of the relevant recent studies, such as those from Refs. 25-30,33,39-42,44,46, is the translational symmetry of the media in the direction parallel to the resonators' long axis. However, the expanding availability and use of laterally confined YIG magnonic waveguides[54,56] urges their implementation with chiral magnonic resonators, e.g. as in Refs. 15,16. This also implies a need to miniaturise the resonators themselves.

**B. Enhancement and tuning of the chiral magneto-dipole coupling**

We have seen that the coupling chirality and its strength relative to the dissipation are of paramount importance for the device design. Among other functionalities, the phase inversion in item A2 is most demanding, requiring $\varGamma_R \gg \varGamma_0 + \varGamma_L$. Hence, the following opportunities, in addition to the use of dark modes in the resonator (A4), should be explored.

B1. *Reduction of the dissipative linewidth.* One way forward is to suppress $\varGamma_0$. Ideally, to cut the power consumption, this should be done via reduction of the Gilbert damping in the resonator. To achieve this, one could use e.g. free-standing YIG structures as in Ref. 58,59. Alternatively, one could explore compensating the damping, e.g. using spin currents[38,45] or parallel pumping.[60]

B2. *Enhancing the coupling chirality by making the precession more circular.* Using the bias field to control the precession ellipticity in the medium, one can suppress the coupling's chirality altogether.[28,29] So, one could also try to control this ellipticity to enhance the coupling. Also,

Equation (1) tells us that the precession ellipticity in the resonator is as important as that in the medium. So, one could use the bias magnetic field or anisotropy (shape[50] or magneto-crystalline[61]) to tilt the resonator's static magnetisation, so as to control the projection of the precession ellipse onto the sagittal plane. One could make use of the propagating mode's angle of incidence[62] and dispersion anisotropy[63,64] to control the orientation of the plane of polarisation of its stray field relative to the resonator's magnetisation. Ultimately, one could use the perpendicular magnetic anisotropy[37,65,66] to control the precession ellipticity directly. One should also pay attention to the switching of the precession ellipticity when the spin wave character changes from propagating (Larmor precession) to evanescent (anti-Larmor precession).[67]

B3. *Enhancing the coupling strength and chirality by slowing down spin waves.* The coupling strength may be enhanced by decreasing the group velocity $v$ in the denominator of $\varGamma_{R(L)} \equiv |\varDelta_{R(L)}|^2/(2v)$, which could be done (with varying efficiency) using the bias magnetic field, applied globally or locally, or using other method of graded index magnonics.[68] One could also make use of that, in fact, $v$ in this expression should be replaced with $v_{R(L)}$ – the group velocity of the right (left) going waves. This could enable one to enhance the radiative linewidths' chirality by non-reciprocally slowing done the propagating modes, e.g. using DMI,[4-8] thickness-asymmetry in the DESW geometry[10-14] or electric field.[69-71] One should note however that the reduced group velocity can slow down the device overall.

B4. *Nonlinearity.* In the context of chiral magnonic resonators, spin wave nonlinearity has hardly been explored at all, although being of ultimate importance for their potential practical applications, e.g. in neuromorphic computing. Fig. 7 shows that the chiral magnonic resonators from Ref. 25 have a nonlinear transmission coefficient mimicking a neuron's activation function. Arrays of such resonators could therefore be used to build magnonic artificial neural networks and reservoir computers.[72,73] This nonlinearity originates from the concentration of the incident spin wave energy in the resonator (Fig. 6 [16,25]) and is stronger than that achieved using e.g. nonlinear interference.[72] The arrays of nonlinear chiral magnonic resonators could still have a smaller footprint than other nonlinear magnonic systems.[73-75]

## C. Towards 2D and 3D chiral architectures

One of the most useful features of the chiral device shown in Fig. 4 is the obvious simplicity of its construction and therefore robustness of its functionality. However, further rich opportunities, of fundamental and (eventually) practical value, will result from increasing the design complexity, limited only by one's imagination.

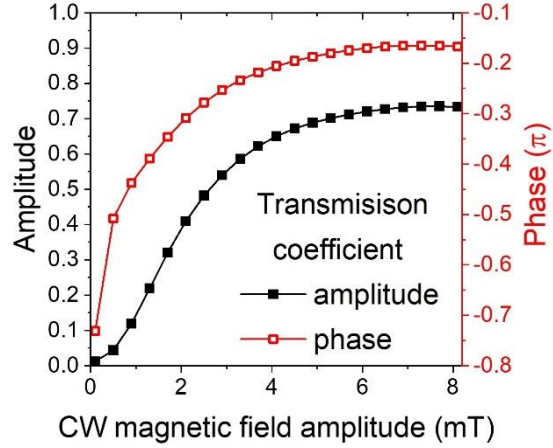

Fig. 7  Numerically simulated spin wave transmission coefficient of a chiral magnonic resonator mimicking a neuron's activation function. The resonator and waveguide are identical to those from Ref. 25 at 15 nm spacing between them. The spin wave frequency is tuned to the CMR's linear transmission minimum. Both amplitude and phase of the SW transmission coefficient strongly vary as a function of the incident SW amplitude: the spin wave neuron "fires".

C1. *Patterned waveguides*. The use of patterned waveguides could be used to control the propagating modes (and therefore their coupling to adjacent chiral magnonic resonators). For instance, the shape anisotropy of the constituent magnetic elements could be used to produce self-biased waveguides magnetised at an angle to the propagation direction,[47,50] influencing the coupling chirality and strength as described in items B2 and B3. Alternatively, the magnonic band structure[76] could be used to explore effects associated with detuning of the resonator's frequency relative to the band gap.[77]

C2. *Finite resonators on continuous media*. One does not need to stick to one dimensional (1D) designs, such as finite resonators on 1D stripe waveguides[15,16] and infinitely long resonators on thin film media. Instead, one could form 2D systems (which could be periodic[41,78-80] or irregular) of finite resonators e.g. on continuous thin films of YIG (Fig. 8), thereby avoiding loss due to edge roughness, and master diffraction as a means of spin wave control.[81,82] As an example, artificial spin ices[78-80] on YIG would represent a promising object for investigation.

C3. *Chiral magnonic crystals*. Periodic arrays of chiral magnonic resonators formed on continuous or patterned waveguides represent chiral magnonic crystals, the physics of which just started to be explored.[83,84]

C4. *Chirality in 2D patterned structures*. Imagine that Figs. 1 (b-e) showed top views of magnonic structures patterned on substrates parallel to the $(x, z)$ plane and having finite thickness in the $y$ direction. Provided the thickness of such side-flipped structures was large enough, the magneto-dipole coupling between the resonators and waveguides would be both strong and chiral. Furthermore, the waveguide could be patterned, as in item C1. Then, considering a cylindrical resonator and a waveguide formed by an array of identical but oppositely magnetised cylinders, we would arrive to the system considered in Ref. 85. In the simulations, the system showed behaviour similar to that exemplified in Fig. 3 and was used to build a magnonic XOR gate.[85] Overall, such 2D patterned chiral structures could be simpler to fabricate in comparison to some other systems considered here, yet exhibiting the same physics.

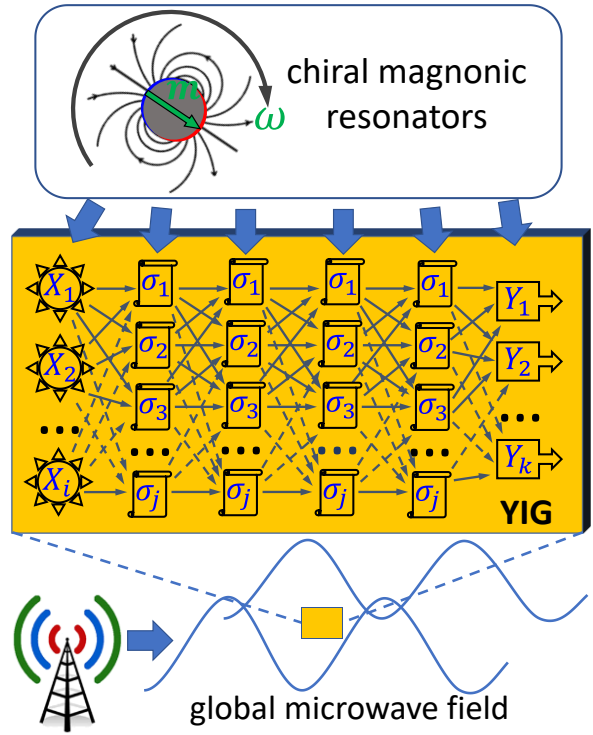

Fig. 8  A visionary schematic of a versatile magnonic chip, here having form of a magnonic neural network. Chiral magnonic resonators, formed atop of a magnonic medium, serve as spin wave input couplers ($X_i$), control elements ($\sigma_j$), and output couplers ($Y_k$). The resonators' designs depend on their functions. The input couplers (input layer neurons), $X_i$, exploit resonators' quasi-uniform modes inductively coupled to the global (on the chip's length scales) microwave clock field. The spin wave control elements (deep layers' neurons), $\sigma_j$, are based on the resonators' dark modes, either decoupled from or perhaps weakly coupled to the clock field. The output couplers (output layer's neurons), $Y_k$, may be based on either quasi-uniform or dark modes, depending on the exact coupling mechanism. The magnonic medium could be either continuous or represent a mesh / combination of magnonic waveguides, possibly extending to multiple layers or forming 3D meshes. Magnetic states of the input ($X_i$), control ($\sigma_j$) and output ($Y_k$) resonators can all serve as logic input variables, when used a field programmable gate array. Reproduced (with modifications) from Ref. 16, with permission of AIP Publishing.

C5. *Nonuniform magnetic textures*. Let us consider a long chiral magnonic resonator (e.g. above a YIG film or stripe waveguide) with a 180° domain wall near the middle. Obviously, only one of its two domains would couple to incident spin waves. Absorption of the spin wave in the domain could move the domain wall,[66,86,87] suggesting a memristor action. Alternatively, one could apply an external stimulus to alter the domain wall's position[87-89] and thereby to control the spin wave scattering. One could extend this idea to multiple domain walls (e.g. in a race track [90]), study resonant spin wave scattering from domain walls (rather than domains), and eventually, exploit other magnetic textures[91] in continuous or patterned structures adjacent to the YIG medium.

C6. *Substantially 2D and 3D architectures.* There is a strong potential, both for discoveries and applications, in systems consisting of meshes of magnonic waveguides and multiple resonators. One example is the magnonic holographic memory device,[92,93] representing a mesh of interconnected magnonic waveguides in the nodes of which chiral magnonic resonators could be placed. The re-programmability of the resonators by switching their magnetisations[15,16] renders this a spin wave version of the field programmable gate array (FPGA). Another example is given by waveguides running in close proximity to each other. The waveguides could be coupled to the same chiral magnonic resonator, forming a circulator.[94] Alternatively, with an appropriately graded design,[34,68] one waveguide could serve as a resonator for the other, enabling spin wave steering in arbitrary (e.g. orthogonal) directions,[95] either in plane or normal to the substrate plane, circumventing the need for waveguide bends and junctions. This would represent a crucial step towards truly 3D magnonic architectures.[21] Importantly, as proposed in Ref. 16, such 2D and 3D systems, as well as those in items C2, C4 or C5, could benefit from the resonators' ability to serve as local magnonic sources (Fig. 8), simplifying the design, bringing the power where it is needed, and avoiding unnecessary Joule losses. Such architectures, enabling nonlinear (Figs. 7) mapping between multiple inputs and outputs (Figs. 8), would be essential for creation of magnonic devices for in-memory and neuromorphic computing.[72,73,96,97]

**D. Programming and active control of chiral magnonic resonators**

Practical application of chiral magnonic resonators requires that their function be efficiently controlled. To this end, there are several promising opportunities worthwhile exploring.

D1. *Micromagnetic programmability*. As discussed before, the resonators are naturally programmable through switching the orientation of their (average) magnetisation,[15,16] and one could also explore active control of their micromagnetic states (item C5). The physics of the magnetisation switching and magnetic recording is a well explored and powerful area of magnetism.[38,65,90,98] Its knowledge and methods could be directly applied to programming of chiral magnonic resonators, which are, in this respect, just soft magnetic elements. This does need to be realised in practical resonant devices.

D2. *Bias magnetic and electric fields, spin current.* These could be used to control the propagating modes' dispersion, resonators' modes' frequencies and dissipative linewidth, and the coupling strength and chirality, as discussed in items B1-B3. All this could be done dynamically, at frequencies below the system's spectrum of magnetic excitations. In the same fashion, one could use elastic strain within the concept of magnon straintronics.[99,100] The use of strong modulation could provide the much-needed nonlinearity.[59,101]

D3. *Superconducting spacers, overlayers and substrates.* The use of superconductors (e.g. Nb [102,103]) as spacer material between waveguides and resonators would increase nonlinearity of the coupling (through the Meissner effect) and could enable switching the coupling on and off (through heating or cooling the spacer across the critical temperature). The use of superconducting overlayers or substrates (which can be high-temperature superconductors [104,105]) could enable additional control of the propagating modes' dispersion[102-105] and therefore their coupling to the resonators.

D4. *Spin wave assisted switching.* An intriguing opportunity lies in using the power absorbed by the resonators to assist their switching, e.g. by a magnetic field or spin current. As discussed above, this resonant absorption is not only chiral but also highly sensitive to the angle of spin wave incidence. This could help switch resonators selectively. Analogously to the absorption induced domain wall motion hypothesised in item C5, this would provide another nonlinear feedback mechanism, of great use for in-memory computing and if training arrays of chiral magnonic resonators (e.g. like those shown in Fig. 8) as artificial neural networks.

**VI. Conclusions and outlook**

The key premise of magnonics is its ability to produce a useful practical device or devices, which would either be able to compete with or to complement those currently based on semiconductor technology, or would occupy a completely new niche, such as in-memory or neuromorphic computing. In the discussion above, we have argued that devices based on chiral magnonic resonators are best positioned to achieve exactly that. At the same time, many of the outlined ideas could find application beyond remits of ferromagnet-based magnonics, primarily targeted here. This includes such important areas as antiferromagnonic magnonics,[106] spin-caloritronics,[107] cavity magnonics,[108] and magneto-acoustics,[109] to list a few. Addressing those opportunities in any detail would go well beyond the scope of this short perspective article, and so, we leave them to more distant future.

**Acknowledgements**

This article describes vision that has formed as a result of teamwork at the University of Exeter spanning more than a decade. However, very special thanks go to Y. Au, K. G. Fripp, C. S. Davies, O. Dmytriiev, and F. B. Mushenok for their hard and creative work on the ground and to A. V. Shytov, V. D. Poimanov, O. Kyriienko, and A. N. Kuchko for their insightful comments and illuminating discussions. I am also obliged to K. G. Fripp for his simulations results from which are used for Fig. 7. The research leading to these results has received funding from the Engineering and Physical Sciences Research Council of the United Kingdom under projects EP/L019876/1 and EP/T016574/1, from the European Union's Seventh Framework Programme under grant agreements 228673 (MAGNONICS), 233552 (DYNAMAG), and 247556 (NoWaPhen), and from Horizon 2020 research and innovation program under Marie Skłodowska-Curie grant agreement 644348 (MagIC).

**Conflict of interest**

The authors have no conflicts to disclose.

**Data availability**

Data sharing is not applicable to this article as no new data were created or analysed in this study.